\documentclass[12pt]{article}
\usepackage{graphicx}
\usepackage{bm}
\usepackage{latexsym}
\newcommand{\text}{\mbox}
\newcommand{\beq}{\begin{equation}}
\newcommand{\eeq}{\end{equation}}
\newcommand{\bc}{\begin{center}}
\newcommand{\ec}{\end{center}}
\newcommand{\eeqa}{\end{eqnarray}}
\newcommand{\beqa}{\begin{eqnarray}}

\newcommand{\ra}{\rightarrow}

\newcommand{\ga}{\gamma}

\newcommand{\de}{\delta}

\newcommand{\si}{\sigma}
\newcommand{\ta}{\tau}

\newcommand{\ph}{\phi}

\newcommand{\ed}{\end{document} }
\begin{document}
\raggedright
\parindent3em
\baselineskip=20pt
\title{Intense ultrashort electromagnetic pulses and the equation of motion}

\author{Richard Hammond\thanks{rhammond@email.unc.edu}\\
University of North Carolina\\
Chapel Hill, North Carolina
}
\date{\today}
\maketitle
\begin{abstract}
{The equations of motion of charged particles under the influence of short electromagnetic pulses are investigated. The subcycle regime is considered, and the delta function approximation is applied. The effects of the self force are also considered, and the threshold where radiation becomes important is discussed. A dimensionless parameter is defined that signals the onset of radiation reaction effects.}

\end{abstract}

Very short and even subcycle optical pulses have been gaining increasing 
attention in recent years.\cite{kaplan} The theory of the interaction of short pulses with charged particles has been studied in \cite{hojo}, \cite{cheng} in one dimension. 
In three dimensions the subcycle problem becomes more complicated, and contentious \cite{rau}-\cite{wang}, and it has also been studied in plasmas.\cite{hazeltine} Often, as the pulses decrease in their temporal span, the intensity rises correspondingly. In fact, intensities of $10^{22}$
W cm$^{-2}$ have been reached, and this number is expected to go even higher.\cite{michigan} At such extreme conditions, the radiation reaction force should be examined, and we will examine the intensities and pulse durations where the onset of the self force becomes important.

To begin, a nonrelativistic approximation is used to assess the use of a delta function to model a short pulse. It is shown to be in agreement with the exact solution of a Gaussian pulse, the approximation improving as the Gaussian pulse becomes smaller. It is then shown how the delta approximation works for the relativistic case, and finally, self force effects are considered.

For a short pulse the slowly varying envelope approximation fails, and for a subcycle pulse the whole notion of a wave train is derailed. Opposite the limit of a monochromatic wave sits a delta function, and in the following we examine the usefulness of this, other extreme, approximation. The value of this approximation lies in the following observation. As the pulse becomes ever smaller, the exact pulse shape may not be known exactly, and furthermore, sometimes the detailed motion during the fleeting moment of interaction is not of interest anyway, where the final velocity and energy are of more interest.

The most drastic approximation to be made is the one dimensional model, meaning that the pulse, or waveform is of the form $f(z-ct)$, i.e., the plane wave approximation (although not necessarily a wave). Real pulses are focused down in a variety of ways and, regardless of how they are made, the field is a function of all of the coordinates. This has been discussed extensively in the references given above. Nevertheless, there are times when the 1-D approximation is useful and sheds light on the more exact physics. This is one of those cases, where we are able to examine both the delta function approximation and self force effects with relative ease.

 For example, consider Gaussian pulse. Non-dimensionalized units are used 
where $z/L \ra z$ and $c t/L \ra t$, where $L$ is taken to be of the order of the wavelength of visible light (for numerical calculations below $L=5,000\AA$), and the field is polarized in the $x$ direction.

\beq\label{egaussian}
{\bm E} =E e^{-((z-t)/w)^2}\bm{ \hat x}
\eeq
where $w$ is a dimensionless parameter fixing the width of the pulse.
This must be accompanied by a magnetic field

\beq\label{bgaussian}
{\bm B} =E e^{-((z-t)/w)^2}\bm{ \hat y}.
\eeq
 For nonrelativistic
dynamics one may ignore the magnetic field and integrate ${\bm F}=m{\bm a}=q{\bm E}$, for a particle with charge $q$ and mass $m$. It is helpful to define the impulse
by

\beq
 I=\int_{-\infty}^\infty F dt =qEwL\sqrt{\pi}/c
.\eeq
With this, integrating the equation of motion yields

\beq
{mv\over I}=\frac12 \left(1+\mbox{erf}({t-z\over w})\right)
\eeq
This is plotted in Fig. 1 at $z=0$ for different values of $w$.

As $w$ gets small, we may think of the force on the particle as impulsive, and in fact,
approximate the Gaussian by a delta function\footnote{Since this is a function of $z-t$ it satisfies the wave equation, which also follows from the fact that $\de({z-t\over w})_{,z}=-\de({z-t\over w})_{,t}$} ${\bm E} =N\de\left({z-t\over w}\right)\bm {\hat x}$ where normalization $N$ is fixed by assuming

\beq
\int Ee^{-(z-t)/w)^2}dt=\int N\de\left({z-t\over w}\right)dt
\eeq
which gives $ N=\sqrt{\pi}E$. This insures that the impulse is the same in each case.
The velocity is easily found. Assuming that the initial velocity is zero gives

\beq
v=\frac I m \theta(t)
.\eeq

\begin{figure}[!h]
\centering
\includegraphics[width=7cm]{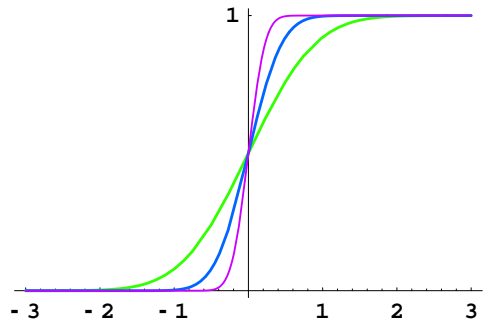}
\caption{$mv/I$ vs $t$ for $w=1,1/2,1/4$, showing that the smaller $w$ is, the more steplike is the response.}
\label{f1}
\end{figure}

This shows that replacing the Gaussian by a delta function reproduces the solution of the equation of motion in the limit of a narrow pulse. In the limit that $w \ra 0$ these results show that the Dirac approximation precisely coincides with the exact result given by the Gaussian approximation. In reality, short pulses have a complicated profile, but these results show the advantage of the delta function approach: It changes the problem from solving a differential equation to an algebraic equation.

To consider the relativistic situation consider the equation of motion

\beq\label{rel}
{dv^\si\over d\ta}={q\over mc}F^{\si\nu}v_\nu
.\eeq
We consider an electromagnetic wave of the form  

\beq\label{fx}
{\bm E} =Ef(z-t)\bm{ \hat x}
\eeq
and

\beq\label{fy}
{\bm B} =Ef(z-t)\bm{ \hat y}.
\eeq
This represents a plane wave of amplitude $E$, polarized in the $x$ direction, described by any dimensionless function $f$. With this (\ref{rel}) gives,

\beq\label{v0}
{dv^0\over d\ta}=afv^1
\eeq

\beq\label{v1orig}
{dv^1\over d\ta}=af(v^0-v^3)
\eeq

\beq\label{v2}
{dv^2\over d\ta}=0
\eeq

\beq\label{v3}
{dv^3\over d\ta}=afv^1
\eeq
where the dimensionless parameter $a=eLE/(mc^2)$.

As these equations stand, the delta function approximation fails. This is because, since the velocity is like a step function function, the integral of (\ref{v3}), for example, is difficult to assess.\footnote
{For example, $\int\theta(t)\de(t) =1/2$. Since $v^1$ is not exactly a step function, this result is inapplicable as well.}
A better way to proceed is to note that (\ref{v0}) and (\ref{v3}) imply,

\beq\label{v03}
v^0=1+v^3
\eeq
which leaves the pair

\beq\label{eq2}
{dv^0\over d\ta}=afv^1
\eeq

\beq\label{v1}
{dv^1\over d\ta}=af
.\eeq
These imply, 

\beq\label{v01}
v^0=1+(v^1)^2/2
.\eeq
Using the integral of (\ref{v03}) in the right hand side of (\ref{v1}) yields

\beq\label{v1f}
v^1=a\int f(-\tau) d\ta
,\eeq
a fascinating result. It states that the $x$ component of the four velocity is essentially equal to the nonrelativistic three velocity evaluated at the proper time $\ta$. With this,
(\ref{v03}), and (\ref{v01}), the relativistic solution is completely determined in terms of the nonrelativistic solution for an arbitray 1D wave form $f(z-t)$.

\begin{figure}[!h]
\centering
\includegraphics[width=10cm]{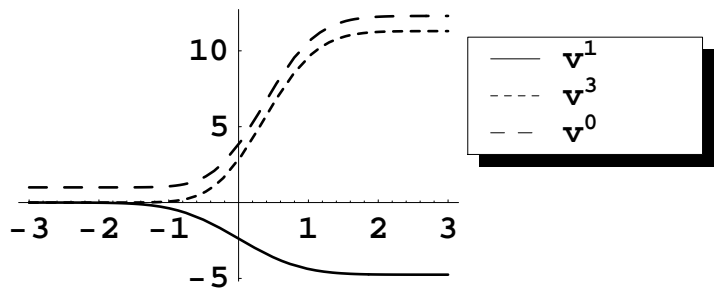}
\caption{The asymptotic regions agress with the delta function approximation, for $v$ vs. $t$.}
\label{3vs}
\end{figure}

For example, the delta function approximation may be used in (\ref{v1f}). With the integral of (\ref{v03}), and letting $f\ra \sqrt{\pi} \de$, we find 
\beq\label{vstep}
v^1=a\sqrt{\pi}\theta(\ta)\ \ \ \ v^3=\pi a^2/2\theta(\ta)\ \ \ \ v^0=1+\pi a^2/2\theta(\ta)
.\eeq
It is easy to see that this agrees with the asymptotic form of the analytical solution,

\beq\label{v1exact}
v^1(\ta)={a\sqrt{\pi}\over2}(1+\mbox{erf}(\ta))
,\eeq
and the other components are found by simple alegra.
For example, for an intensity of $10^{18}$  W cm$^{-2}$, which corresponds to $a=-2.7$, the exact analytical solutions are plotted in
Fig. \ref{3vs}. The asymptotic region is seen to agree with the delta function approximation.

Another useful observation from (\ref{vstep}) is that for high intensity, $v^3\approx v^0$. In particular this is valid for $a>>1$. For example, at $I=10^{19}$  W cm$^{-2}$ the difference between  $v^3$ and $v^0$ is less than one percent.

\begin{figure}[!h]
\centering
\includegraphics[width=10cm]{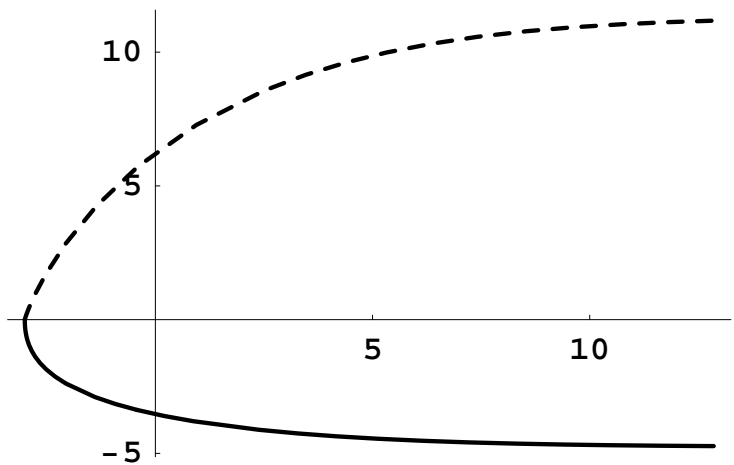}
\caption{The $x$ and $z$ components of the four velocity versus $t$.}
\label{pp}
\end{figure}

One may also solve (\ref{v01}) with (\ref{v1exact}) for the time $t$ in terms of the proper time $\ta$,

\beq
t=\frac{1}{8} a^2 \pi  \left(\tau  \mbox{erf}(\tau )^2+2
   \left(\tau +\frac{e^{-\tau ^2}}{\sqrt{\pi }}\right)
   \mbox{erf}(\tau )+\tau -\sqrt{\frac{2}{\pi }}
   \mbox{erf}\left(\sqrt{2} \tau \right)+\frac{2 e^{-\tau
   ^2}}{\sqrt{\pi }}\right)
,\eeq
which is used in Fig. \ref{pp}.

Having an analytical expression for the velocity is useful for looking at the self force. The equation of motion with radiation reaction forces is
\footnote{We assume $v_\si v^\si=1$. In the literature, some take $v_\si v^\si=-1$, which changes signs in the self force.}

\begin{equation}\label{lad1}
{dv^\sigma\over d\tau}=af^{\si\mu}v_\mu+b(\ddot v^\si+\dot v^\nu\dot v_\nu v^\si)
\end{equation}
where  the dimensionless parameter $b\equiv c\ta_0/L$, $\ta_0=2q^2/(3mc^3)$ and the overdots imply differentiation with respect to the proper time. This is called the Lorentz Abraham Dirac equation, but the other form of the equation of motion is obtained
by starting with (\ref{lad1}) but using 

\beq
{dv^\si\over d\tau}=af^{\si\mu}v_\mu
\eeq
in the terms multiplied by $b$, which leads to,

\beq\label{LLR}
{dv^\si\over d\tau}= af^{\si\mu}v_\mu+  ab\dot f^{\si\mu}v_\mu
+a^2b(f^{\si\ga}f_\ga^{\ \ph}v_\ph+f^{\nu\ga}v_\ga f_\nu^{\ \ph}v_\ph v^\si)
,\eeq
which is called the Landau Lifshitz Rohrlich equation. The LAD equation fell under bad times to the the runaway solutions, preaceleration issues, or the apparent need to invoke non-zero size particles,\cite{ori} while the LLR equations seems to be only an approximation. For a nice introduction to the issues, with historical notes, one may consult Rohrlich\cite{rohrlich}, which has many of the seminal references and discusses the distinction between ``self forces'' and ``radiation reaction forces.'' More recent work considers the problem in various dimensions,\cite{galtsov}\cite{kazinski} the effect of the magnetic dipole,\cite{meter}\cite{heras}, connections to QED\cite{higuchi} and vacuum polarization,\cite{binder}, mass conversion\cite{bosanac}, and hydrogenic orbits.\cite{cole}

Whenever $b$ is small, which is true for all but the most extreme light, it is sensible to expand the solution in terms of this parameter,

\beq
v^\si={_0v}^\si+{_1v}^\si+...
\eeq
where ${_0v}^\si$ is the solution without self forces, ${_1v}^\si$ is ${\cal O}(b)$, and so on. Then, for the plane polarized field used above we have,

\beq\label{v0r}
{d{_1v}^0\over d\ta}=af{_1v}^1+g^0({_0v})+{\cal O}(b^2)
\eeq

\beq\label{v1rr}
{d{_1v}^1\over d\ta}=af({_1v}^0-{_1v}^3)+g^1({_0v})+{\cal O}(b^2)
\eeq

\beq\label{v3rr}
{d{_1v}^3\over d\ta}=af{_1v}^1+g^3({_0v})+{\cal O}(b^2)
.\eeq
where $g^\si$ represents  the self force, and where
${_0v}^\si$ is known, given by (\ref{v1f}), (\ref{v03}), and (\ref{v01}). These equations show that $\dot v^0=\dot v^3+{\cal O}(b^2)$, a very useful result, which allows us to use $v^0-1= v^3$ in the ${\cal O}(b)$ equations. With this, we have,
calling $\ph\equiv {_0}v^1$ and dropping the subscripts, to ${ \cal O}(b)$

\beq\label{v0ob}
{dv^0\over d\ta}=afv^1+ab(\ph\dot f-af^2\ph^2/2)
\eeq
and

\beq\label{v1ob}
{dv^1\over d\ta}=af+ab(\dot f-a\ph f^2)
\eeq

\begin{figure}
\begin{minipage}[t]{8cm}
\includegraphics[width=0.9\textwidth]{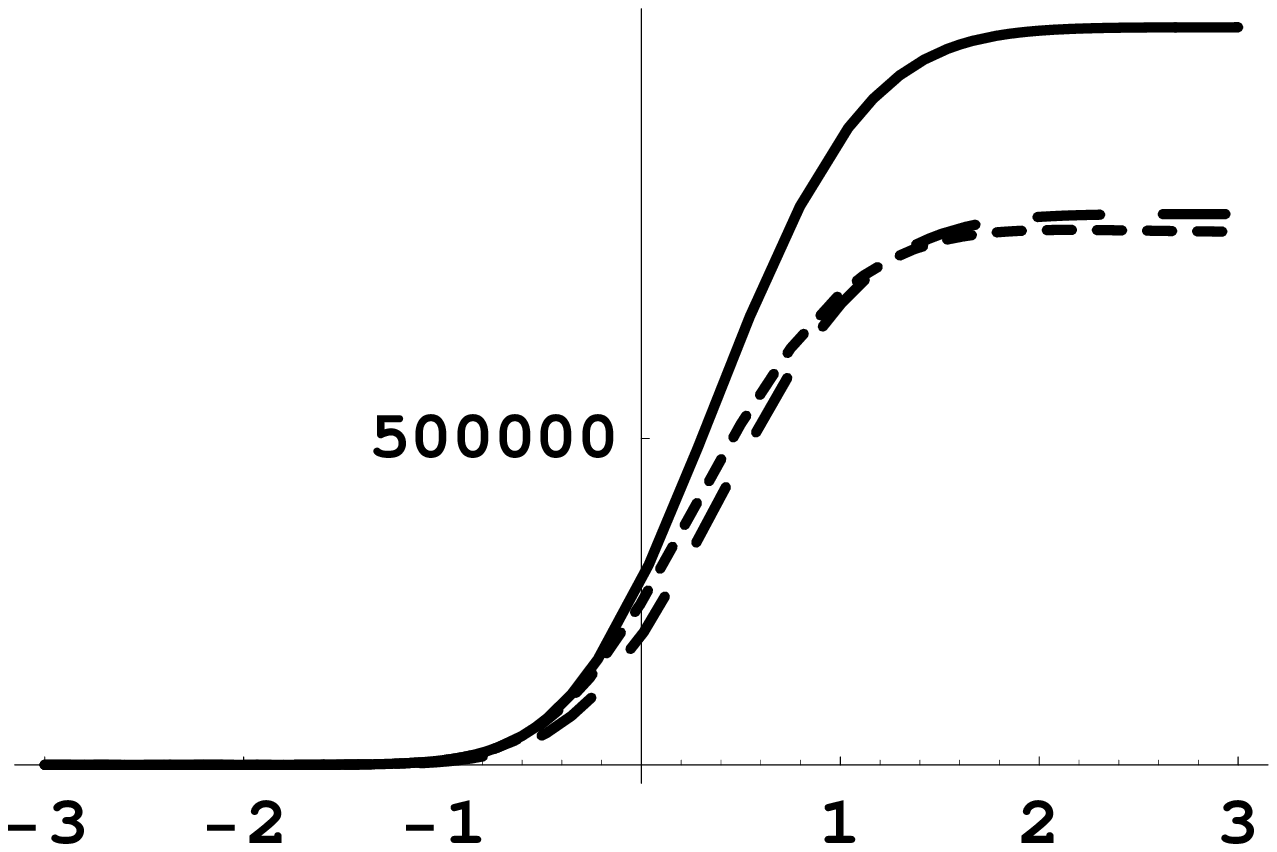}
\caption{$v_0$ plotted against $100t$ with no radiation reaction, as a solution to (\ref{v0ob}) and (\ref{v1ob}), and a numerical solution to the exact equations \ref{LLR}.}
\label{g3vs0}
\end{minipage} 
\hfill
\begin{minipage}[t]{8cm}
\includegraphics[width=0.9\textwidth]{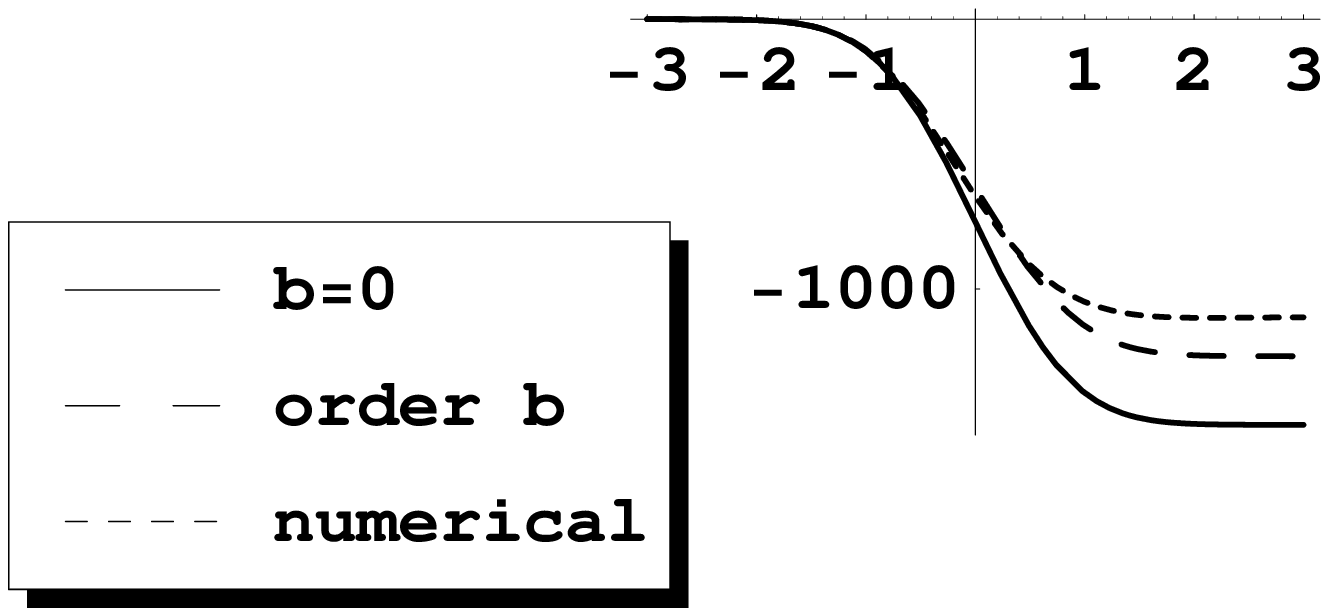}
\caption{$v_1$}
\label{g3vs1}
\end{minipage}
\hfill
\end{figure}

It is noteworthy that these equations are obtained by using {\em either} the LAD or the LLR form for the self force. We use

\beq
f={e^{-((z-t)/w)^2}\over w}\ra {e^{-(\ta/w)^2}\over w}
,\eeq
where the last part holds to ${\cal O}(b^2)$.
One can see that the self forces become important as
$r\equiv a^2b\sqrt{\pi}/w\ra 1$. We examine this, and the accuracy of the approximations, for specificic cases. For an intensity of $10^{23}$W cm$^{-2}$ ($a\sim -850$) and $w=100$, $b\sim 0.5$, the velocities are given in Figs. \ref{g3vs0} and \ref{g3vs1}. It is seen explicitly, as we expect, the effect of the radiation reaction is to reduce the final velocity of the particle. If the intensity is increased by one order of magnitude, so does $r$, and the results diverge drastically, as expected.\footnote
{As a partial check on the numerical work, which was accomplished using Mathematica, $v_\si v^\si$ was plotted, a result not used explicitly in the calculations. The value was always within a tenth of one percent of unity.}

\begin{figure}
\begin{minipage}[t]{8cm}
\includegraphics[width=0.9\textwidth]{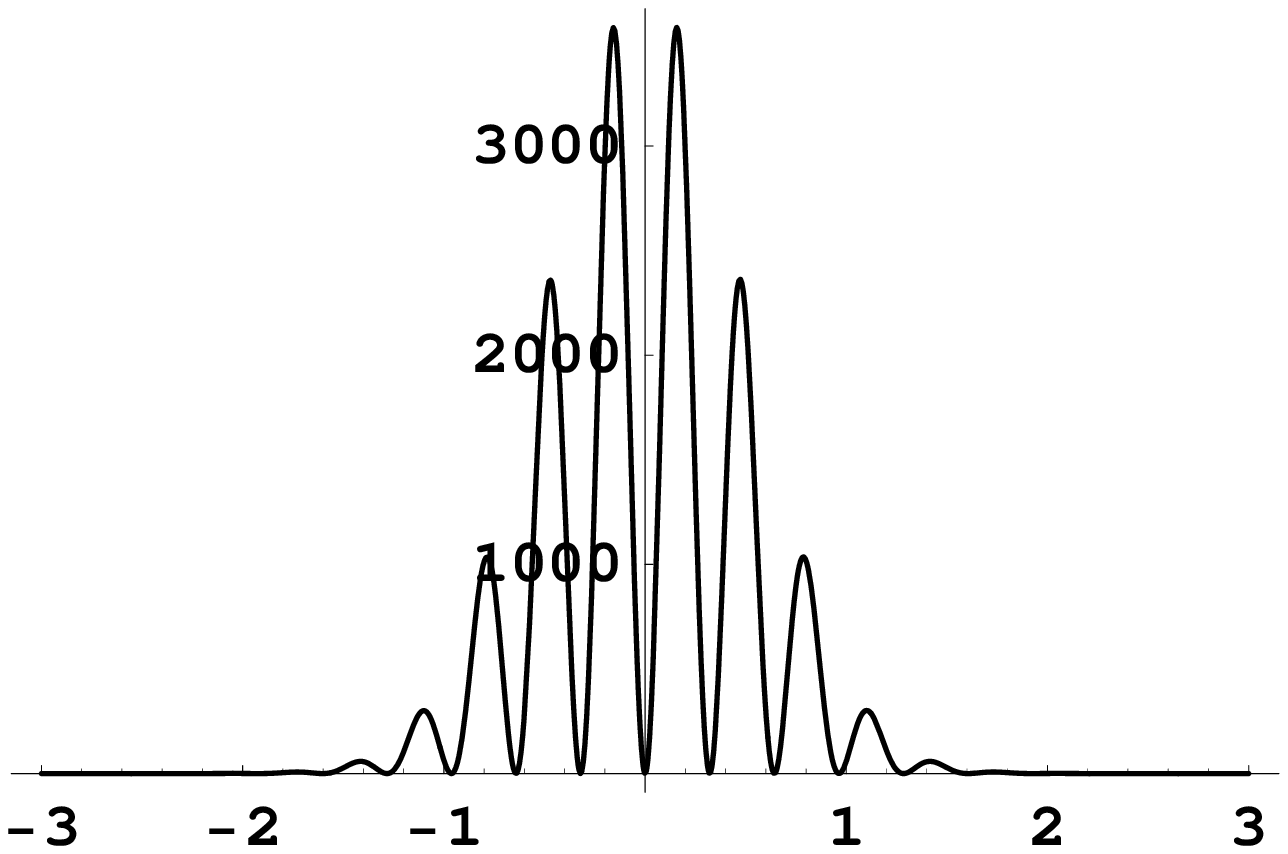}
\caption{$v_z$ plotted against the proper time for $I=10^{23}$W cm$^{-2}$.}
\label{proper}
\end{minipage} 
\hfill
\begin{minipage}[t]{8cm}
\includegraphics[width=0.9\textwidth]{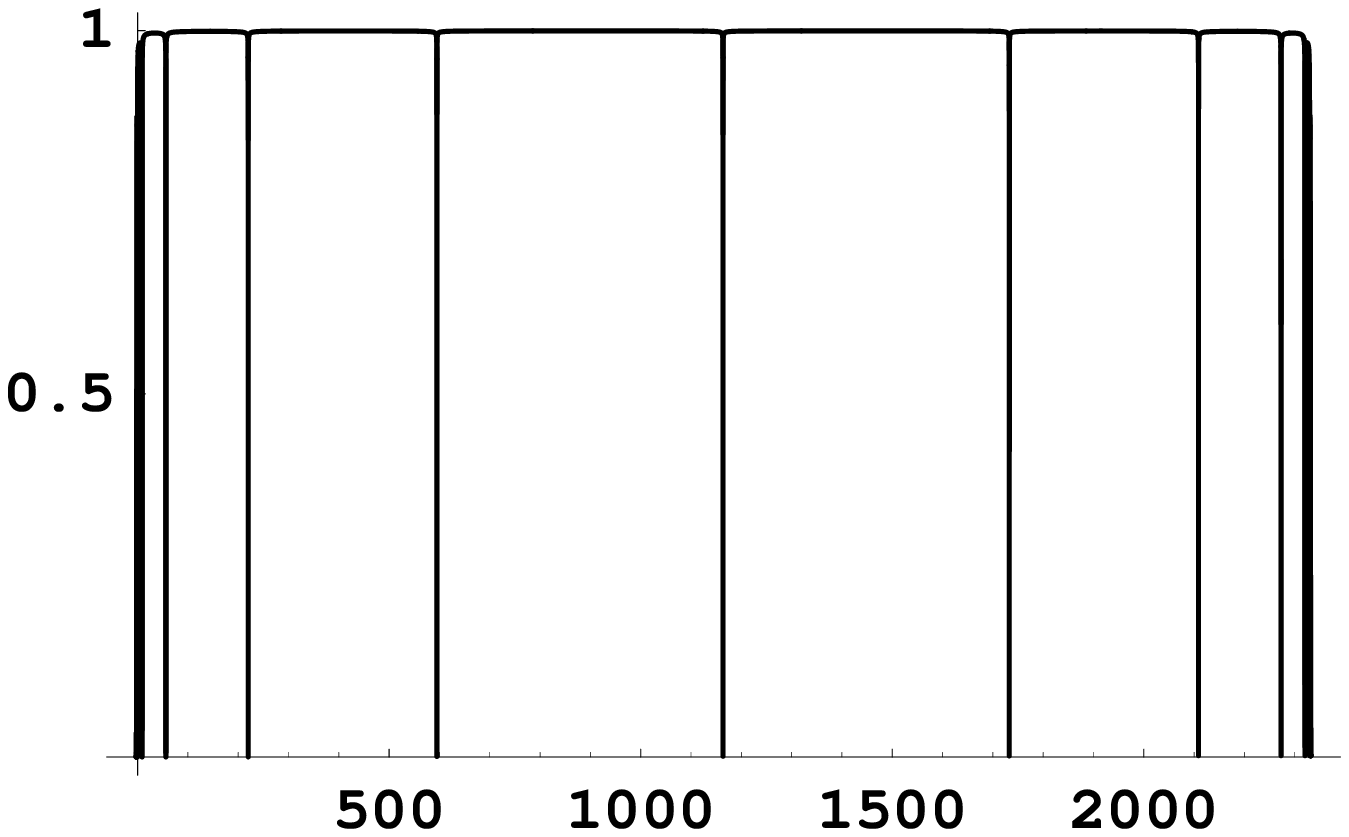}
\caption{$dz/dt$ vs. $t$, corrsponding to Fig. \ref{proper}.}
\label{vvst}
\end{minipage}
\hfill
\end{figure}

As a final example let us consider a pulse of soft x-rays, where we take

\beq
f={e^{-((z-t)/w)^2}\over w}\cos(\Omega(z-t))
\eeq
where the dimensionless $\Omega$ determines the frequency. In this case the delta approximation is invalid, but the expansion in terms of $b$ is still useful. 
To zero order in $b$ we have,

\beq
{_0v}^1=
\frac{1}{4} a e^{-\frac{1}{4} w^2 \Omega ^2} \sqrt{\pi }
   \left(\text{erf}\left(\frac{x}{w}-\frac{i w \Omega
   }{2}\right)+\text{erf}\left(\frac{x}{w}+\frac{i w \Omega
   }{2}\right)\right)
.\eeq
This of course is real, which is seen directly by writing the error function in terms of the imaginary error function. Before considering x-rays it is interesting to consider the case that $\Omega=1=w$ at $10^{23}$W cm$^{-2}$. This is just below the radiation reaction ``threshold,'' but it is interesting to see how strongly relativisitic the solution is. This is evident in Figs. \ref{proper} and \ref{vvst}, which show the four velocity as a function of proper time and the corresponding velocity ($dz/dt$) versus $t$.

\begin{figure}
\begin{minipage}[t]{8cm}
\includegraphics[width=0.9\textwidth]{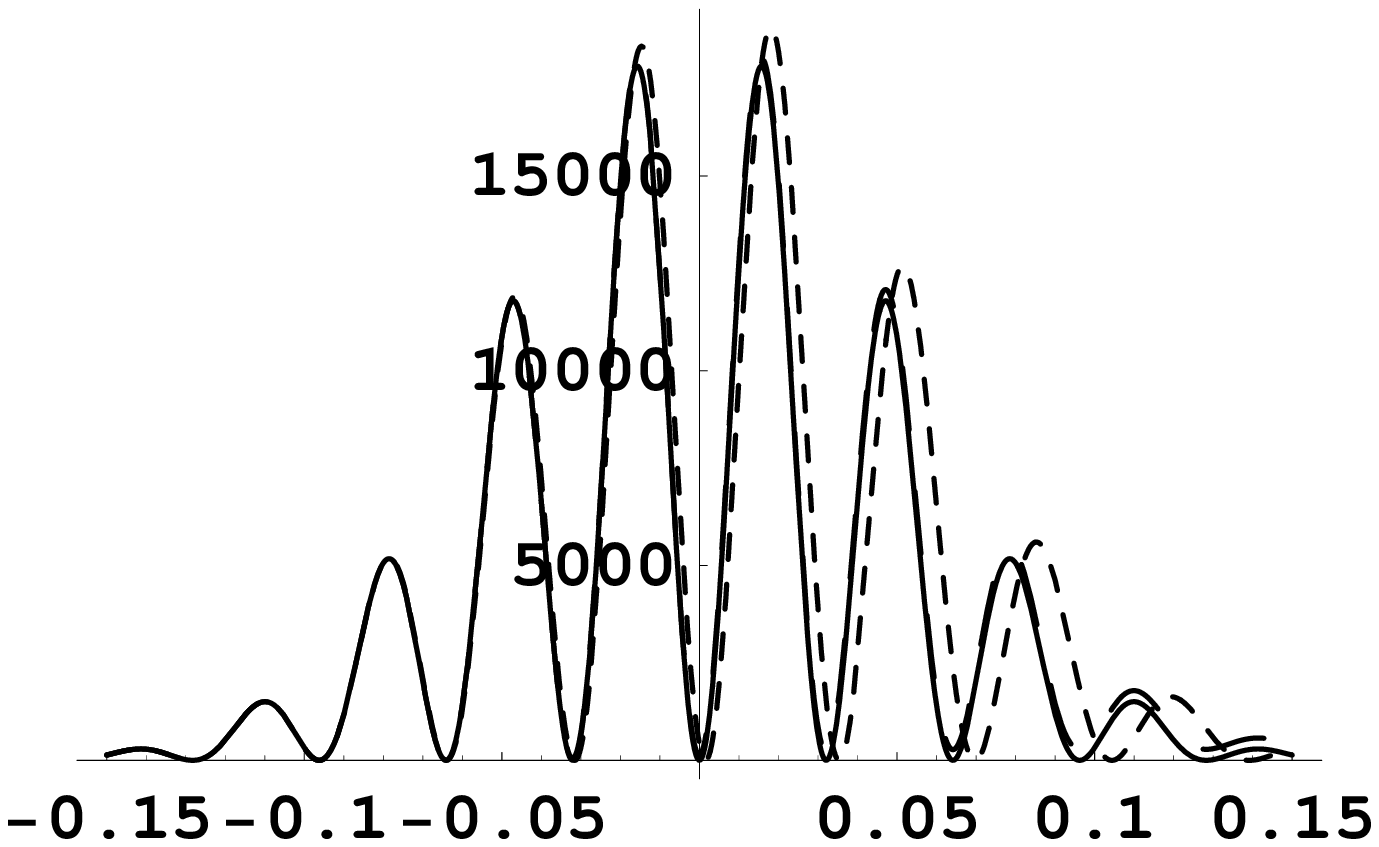}
\caption{$v_0$ plotted against $10t$ with no radiation reaction, as a solution to (\ref{v0ob}) and (\ref{v1ob}), and a numerical solution to the exact equations \ref{LLR}.}
\label{g3vs0cos}
\end{minipage} 
\hfill
\begin{minipage}[t]{8cm}
\includegraphics[width=0.9\textwidth]{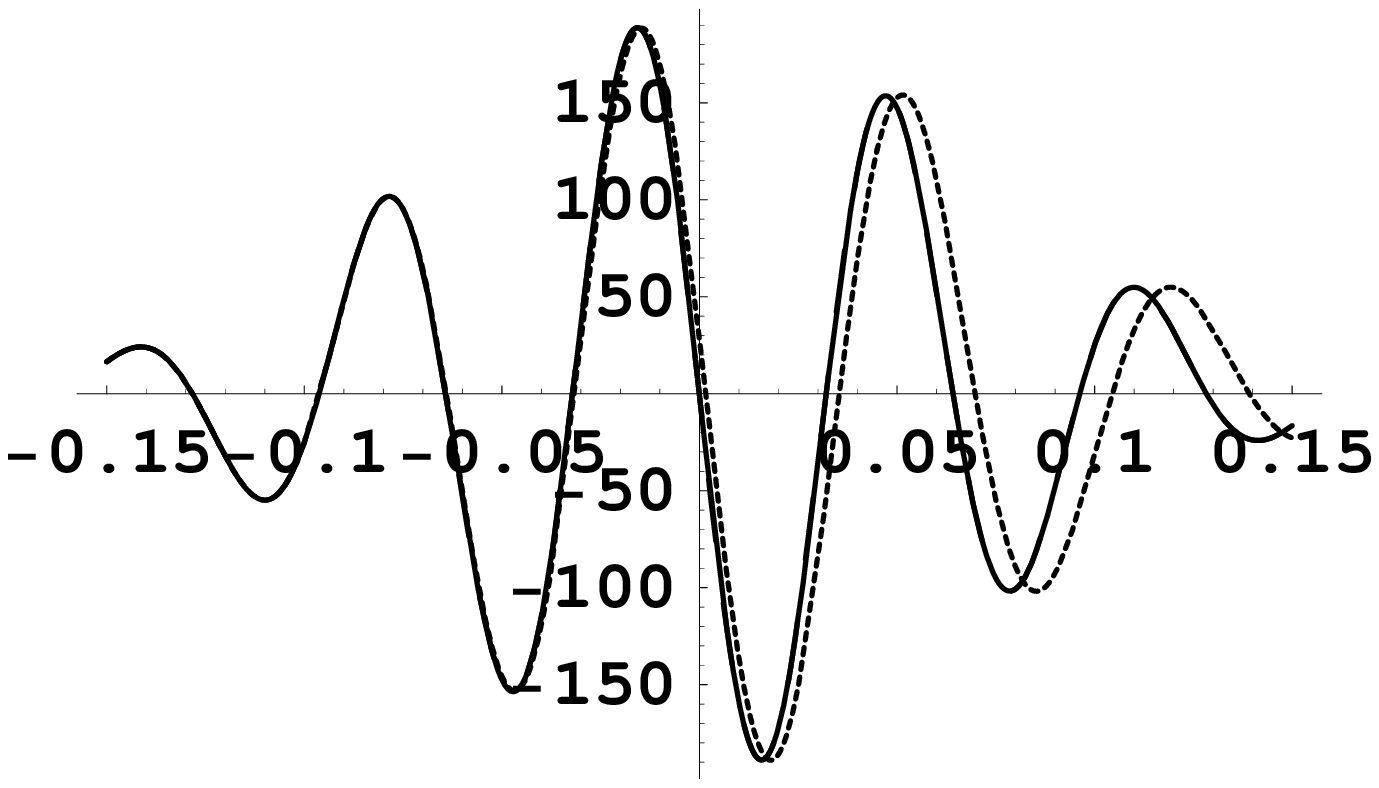}
\caption{$v_1$}
\label{g3vs1cos}
\end{minipage}
\hfill
\end{figure}

For soft x-rays
$\Omega\sim 100$ (50$\AA$ radiation) and we take $w=1/10$ ($\sim 200$as pulse).
The results are plotted in Figs. \ref{g3vs0cos} and \ref{g3vs1cos} for  $I=5\times10^{23}$  W cm$^{-2}$.

In order to assess the vailidty of the approximations we may start with the equations, for any pulse of form considered above,

\beq\label{v0ex}
{dv^0\over d\ta}=a(f+b\dot f)v^1+a^2bf^2(v^0 +v^3-Z^2v^0)
\eeq

\beq\label{v1ex}
{dv^1\over d\ta}=a(f+ b\dot f)Z-ba^2f^2Z^2v^1
\eeq

\beq\label{v3ex}
{dv^3\over d\ta}=a(f+b\dot f)v^1+ba^2f^2(v^0 +v^3-Z^2v^3)
\eeq
where $Z=v^0-v^3$. From (\ref{v0ex}) and (\ref{v3ex}) we have

\beq\label{Z}
Z^{-2}=1+2ba^2\int f^2d\ta
.\eeq
This shows that for $b=0$, $Z=0$, which we found above. Since the integral is bounded, $Z\approx 1$ for small $b$. We can use this to investigate a self consistent solution to this equation by using the $Z=1$ (which implies $t-z=\ta$) in $f$ to find

\beq
Z^{-2}=1+ba^2w\sqrt{{\pi\over2}}\left(1+\mbox{erf}(\sqrt{2}\ta/w)\right)
.\eeq    
This shows that the approximation $Z=1$ is valid as long as $ba^2w\sqrt{{\pi\over2}}<<1$.

In summary, it has been shown that the delta function approximation is applicable for both  the relativistic and non-relativistic case for any incident traveling wave of the form $f(z-t)$. For the relativistic case, it is shown how an exact solution may be found, and these results are used in the investigation of self forces.

\ed